\documentclass[12pt]{article}
\usepackage{epsfig}
\begin{document}
\begin{titlepage}
\begin{center}
{\hbox to\hsize{\hfill UMD-PP-02-035}}

\bigskip
\vspace{4\baselineskip}

\textbf{\Large 
Positively-deflected anomaly mediation 
}\\
\bigskip
\bigskip
\bigskip
\bigskip
\textbf{Nobuchika Okada}\\
\smallskip
\bigskip
\textit{\small
Department of Physics, 
University of Maryland \\
College Park, MD 20742, USA
}

\bigskip

{\tt okadan@physics.umd.edu}

\vspace{4\baselineskip}
\textbf{Abstract}\\
\end{center}
\noindent 
We generalize the so-called 
``deflected anomaly mediation'' scenario to the case 
where threshold corrections of heavy messengers 
to the sparticle squared-masses are positive. 
A concrete model realizing this scenario is also presented. 
The tachyonic slepton problem can be fixed 
with only a pair of messengers. 
The resultant sparticle mass spectrum 
is quite different from that 
in the conventional deflected anomaly mediation scenario, 
but is similar to the one in the gauge mediation scenario.  
The lightest sparticle is mostly B-ino. 

\end{titlepage}

\setcounter{footnote}{0}
\newpage
\section{Introduction}

Supersymmetry (SUSY) extension is one of 
the most promising way to solve 
the gauge hierarchy problem in the standard model. 
However, since none of the sparticles 
have been observed yet, supersymmetry must be broken 
at low energies. 
In addition, sparticle masses are severely 
constrained by experiments, 
since arbitrary soft supersymmetry breaking masses 
cause too large flavor changing neutral currents 
(SUSY flavor problem). 
Finding a simple mechanism of supersymmetry breaking 
and its mediation is one of the most important task 
for realistic supersymmetric theories. 

The anomaly mediated supersymmetry breaking (AMSB) 
\cite{Randall:1998uk} \cite{Giudice:1998xp}
is one of the most attractive scenario in supergravity. 
This is because it predicts 
the sparticle mass spectrum being flavor blind 
and thus solves the SUSY flavor problem 
automatically. 
In addition, since SUSY breaking is mediated through 
the superconformal anomaly, sparticle masses at low energies 
are insensitive to any high energy theories 
and mechanism of SUSY breaking, 
namely, model independent. 

In order to realize the AMSB scenario, 
the sequestering between the visible sector 
and the hidden sector in supergravity action 
is necessary. 
This is naturally realized in the five dimensional brane 
world scenario \cite{Randall:1998uk} \cite{Luty:1999cz}, 
where the visible and hidden sectors are confined 
on the different branes geometrically separated%
\footnote{
It is recently pointed out \cite{Anisimov:2001zz} 
that the sequestering is not a generic prediction 
of the string theories, 
even if two branes are geometrically separated. 
In this paper, we simply assume that 
we are on a special point in the string moduli 
(if the string theory is the ultimate theory behind us),  
where the sequestering is realized. 
},
or in the models where the contact terms between 
the hidden and visible sectors 
are suppressed dynamically by a conformal sector 
\cite{Luty:2001jh}. 
For simplicity, we assume the sequestering in this paper. 
 
Unfortunately, the pure AMSB scenario 
is obviously excluded, since it predicts 
slepton squared-masses being negative. 
There have been many attempts to solve 
this ``tachyonic slepton'' problem 
by taking into account 
additional positive contributions 
to the sparticle squared-masses  
at the tree level 
\cite{Randall:1998uk} \cite{tree} 
or at the quantum level 
\cite{Pomarol:1999ie} \cite{Katz:1999uw} \cite{quantum}. 

One of elegant scenarios is the so-called 
``deflected anomaly mediation'' scenario 
proposed by Pomarol and Rattazzi \cite{Pomarol:1999ie}.  
We introduce the messenger sector 
with $N$ flavors of messengers such that 
\begin{eqnarray}
 W = \sum_{i=1}^N  \lambda_i  S \overline{\Psi}_i \Psi^i \; , 
\label{messenger}
\end{eqnarray} 
where $\overline{\Psi}_i$ and $\Psi^i$ are the messengers 
in $\bf{\overline{5}}+ \bf{5}$ representation 
under the gauge group $SU(5)$%
\footnote{
We use the conventional $SU(5)$ GUT notation 
in this paper. 
In this notation, the beta function coefficient 
$b_1=-33/5$ and the quadratic Casimir $C = 3/5 Y^2$ 
for the $U(1)_Y$ hypercharge. 
}, and $S$ is the singlet superfield. 
If vacuum expectation values of the scalar component ($S$) 
and the F-component ($F_S$) of the singlet superfield 
are generated, 
new contributions to sparticle masses 
develop through the same manner as 
in the gauge mediation scenario \cite{gmsb} \cite{gmsb2}. 
As a result, sparticle masses are deflected 
from the pure AMSB trajectory of 
the renormalization group equations, 
and the tachyonic slepton problem can be fixed. 
In addition, this scenario predicts 
the specific sparticle mass spectrum \cite{Pomarol:1999ie}. 
Furthermore, detailed phenomenology 
was discussed \cite{Rattazzi:1999qg}, 
and the extension to the model with axion 
was proposed \cite{Abe:2001cg}. 

The crucial difference from the gauge mediation scenario 
is that the SUSY breaking in the messenger sector 
is originated from the anomaly mediation. 
Therefore, non-zero F-component 
of the compensating multiplet ($F_\phi$) 
is the unique source of SUSY breaking 
in this scenario. 
This fact allows us to parameterize 
the SUSY breaking order parameter 
in the messenger sector 
such as (see Eq.~(\ref{basic}) for our notation)
\begin{eqnarray}  
 \frac{F_S}{S} = d  F_\phi \;  .
\label{Fterm}
\end{eqnarray}  
Here we introduced the parameter $d$ which characterizes 
how the sparticle masses are deflected 
from the pure AMSB trajectory. 
We call $d$ ``deflection parameter'' in this paper.  
Note that $d$ should be real and, at most, of order one, 
$|d| \leq {\cal O}(1)$, 
since all the quantities accompanied by SUSY breaking 
should be originated from the anomaly mediation. 

In the conventional deflected anomaly mediation scenario, 
only the negative values for the deflection parameter 
$d < 0$ have been taken into account. 
In this paper we generalize the conventional scenario 
to the case with positive deflection parameter.  
We examine, in the next section, 
how non-zero deflection parameter occurs 
based on a simple superpotential, 
and find that the deflection parameter can be positive. 
In Sec.~3, we present a simple concrete model 
which can realize our scenario. 
The sparticle mass spectrum is presented in Sec.~4. 
We will see that our result is quite different from 
that in the conventional deflected anomaly 
mediation scenario, 
but is similar to that of the gauge mediation scenario. 
The lightest sparticle (LSP) is mostly B-ino 
in our scenario. 
We give a conclusion in the last section.

\section{Generalization and positive deflection parameter} 

In order to fix the tachyonic slepton problem, 
a sizable deflection parameter, 
$|d| \sim {\cal O}(1)$, is necessary. 
In general, this case occurs 
when the superfield $S$ is lighter than $F_\phi$ 
in the SUSY limit and the SUSY breaking effect 
plays an essential role to determine 
the potential minimum of $S$. 

There are two typical cases. 
One is that $S$ has no superpotential 
or an extremely flat potential in the SUSY limit, 
and the potential minimum is determined 
essentially by the effective Kahler potential 
including the anomaly mediation. 
The other case is that $S$ has a superpotential, 
which plays an essential role to determine 
the potential minimum 
after the SUSY breaking effects are taken into account. 
In the first case, the deflection parameter 
is found to be $d \sim -1$. 
This is nothing but the case 
mainly discussed in the conventional deflected 
anomaly mediation scenario. 
On the other hand, the deflection parameter 
can generally be positive in the latter case. 
In fact, we can construct a model 
which realizes $ d > 0 $. 

Let us begin with the supergravity Lagrangian 
for $S$ in the superconformal framework 
\cite{Cremmer:1982en} \cite{Kugo:cu} 
(supposing SUSY breaking in the hidden sector and 
fine-tuning of the vanishing cosmological constant), 
\begin{eqnarray} 
{\cal L} = \int  d^4 \theta \;  
\phi^\dagger \phi  \; 
{\cal Z}(S^\dagger, S)  S^\dagger S 
+ \left\{ \int d^2 \theta \; \phi^3 W(S) + h.c. 
\right\} \;  , 
\label{basic} 
\end{eqnarray}  
where ${\cal Z}$ is the supersymmetric wave function 
renormalization coefficient, 
$W$ is the superpotential (except for Eq.~(\ref{messenger})),
and $\phi$ is the chiral compensating multiplet 
expanded as $\phi=1 +\theta^2 F_\phi$ 
with the unique SUSY breaking source $F_\phi$ 
being the same order as the gravitino mass. 
The scalar potential can be read off as 
\begin{eqnarray} 
V= \frac{\partial^2 {\cal K}}{\partial S^\dagger \partial S} 
|F_S|^2 - {\cal K} |F_\phi|^2 
-3 F_\phi W -3 F_\phi^\dagger W^\dagger 
\label{potential}
\end{eqnarray} 
with the auxiliary field given by
\begin{eqnarray} 
  F_S = - \left( 
  \frac{\partial^2 {\cal K}}{\partial S^\dagger \partial S} 
  \right)^{-1} 
  \left( \frac{\partial {\cal K}}{\partial S^\dagger} F_\phi 
  + \frac{\partial W^\dagger}{\partial S^\dagger}  
  \right) \; , 
\label{FSterm}
\end{eqnarray} 
where ${\cal K}={\cal Z}(S^\dagger, S)  S^\dagger S $ 
is the effective ``Kahler potential'' 
in the superconformal framework. 

Let us first consider the case 
where $S$ has no superpotential 
or the superpotential plays no essential role 
to determine the potential minimum. 
In this case, the superpotential term 
in Eq.~(\ref{FSterm}) can be ignored. 
In addition, since ${\cal Z} \sim 1$ 
is usually a very slowly varying function 
of $S^\dagger$ and $S$ in perturbation theory, 
$\partial {\cal Z}/\partial S^\dagger $, 
$\partial {\cal Z}/\partial S$ and   
$\partial^2 {\cal Z}/\partial S^\dagger \partial S $  
can be all neglected. 
As a result, we obtain $F_S/S \sim - F_\phi$, namely, 
the deflection parameter $d \sim -1$ 
independent of $S$ at the potential minimum. 
This is the case discussed 
in the conventional deflected anomaly mediation scenario. 
We arrive at the same result in the case 
where the potential is bounded 
by higher dimensional Kahler terms 
if $S$ is much smaller than the Planck scale 
\cite{Abe:2001cg}. 

Next consider the case that the potential minimum 
is determined through the superpotential. 
We can take the canonical Kahler potential, 
${\cal K} \sim S^\dagger S$, 
as a good approximation, 
and Eqs.~(\ref{potential}) and (\ref{FSterm}) 
are reduced to simple forms. 
Using the stationary condition, $\partial V/\partial S=0$, 
and Eq.~(\ref{FSterm}), we obtain  
\begin{eqnarray} 
 \frac{F_S}{S} \sim - 2 F_\phi  \; 
 \frac{\frac{\partial W}{\partial S}}
 {S \frac{\partial^2 W}{\partial S^\dagger \partial S}} \; .
 \label{FSterm2} 
\end{eqnarray} 
This is a useful formula, from which we can understand 
that $S$ should be light in the SUSY limit 
in order to obtain a sizable deflection parameter 
$|d| \sim {\cal O}(1)$. 

Suppose that $ \langle S \rangle =S_0 \gg F_\phi$ 
and $S$ is much heavier than the gravitino 
in the SUSY limit. 
After taking the supergravity effects into account, 
the vacuum expectation value of $S$, in general, 
shifts from the value in the SUSY limit 
such as $\langle S \rangle \sim S_0 + {\cal O}(F_\phi)$ 
\cite{Hall:iz}. 
In this case, Eq.~(\ref{FSterm2}) can be expanded 
with respect to the small variable $F_\phi/S_0$ 
with the SUSY vacuum condition 
$\partial W/\partial S (S_0) =0$, 
and we can find that 
$F_S/S \sim F_\phi \times {\cal O}(F_\phi/S_0)$,
that is, $d \sim {\cal O}(F_\phi/S_0) \ll 1$. 
This is the one discussed as the decoupling case 
in the literatures 
\cite{Pomarol:1999ie} \cite{Katz:1999uw} \cite{Abe:2001cg}. 
Note that this example also implies a possibility 
that there would be a sizable effect if $S_0 \leq F_\phi$. 
In this case, detailed analysis of higher order corrections 
is necessary \cite{Katz:1999uw}. 

As a simple and interesting example 
which can incorporate the generalization 
of the deflected anomaly mediation scenario, 
let us introduce the superpotential, 
\begin{eqnarray}
W = M^{3-p} S^p \; , 
\label{superpotential}
\end{eqnarray}   
where $M$ is a mass parameter, and 
$p$ is a real parameter. 
From the general formula of Eq.~(\ref{FSterm2}), 
we find 
\begin{eqnarray} 
\frac{F_S}{S} =  \frac{2}{1-p}  \; . 
\end{eqnarray}  
Note that the deflection parameter becomes positive 
for $p < 1$. 
Since the case $ p \geq 3 $, or $d < 0$, 
has been discussed in the conventional scenario, 
we discuss about only the case with 
the positive deflection parameter 
in the following%
\footnote{
For $p>1$, we can find new consistent solution 
for $ 2 < p < 3$, 
which generalize the conventional scenario 
to the region $-2 < d <0$.
}. 

There is an upper bound on 
the deflection parameter. 
To see this, we analyze the potential in detail. 
It is useful to redefine the superfield 
by $ \phi S \rightarrow S$, 
so as to eliminate the compensating multiplet 
in the (canonical) Kahler potential. 
In this notation, Lagrangian is found to be
\begin{eqnarray} 
{\cal L} = \int d^4 \theta \; S^\dagger S 
+\left\{
\int d^2 \theta \phi^{3-p} M^{3-p}  S^p
+h.c.  \right\} \; .
\end{eqnarray}   
Here $M$ has been taken to be real and positive 
by $U(1)_R$ symmetry rotation without loss of generality. 
Changing variables such that $S=r e^{i \Omega/p}$ 
and $F_\phi =|F_\phi| e^{i \omega}$ 
by using only real parameters, 
the scalar potential is found to be 
\begin{eqnarray} 
V= p^2 M^{6-2 p} r^{2 p -2} 
-2 (3-p) |F_\phi| M^{3-p} r^p 
\cos (\Omega+ \omega)  \; . 
\label{potential2}
\end{eqnarray}  
From the minimization conditions, 
$\partial V/\partial \Omega=0$ and 
$\partial^2 V/ \partial \Omega^2 > 0$,  
we obtain the solution $\Omega=- \omega$  
with the assumption $0 < r < \infty$. 
With this solution, the stationary condition 
with respect to $r$ leads to 
\begin{eqnarray} 
r^{p -2} = \frac{3-p}{p (p-1)} 
M^{p-3} |F_\phi|  \; .
\label{solution}
\end{eqnarray}  
We can find a solution in the region $0 < r < \infty $   
only for $p < 0$. 
Thus, the upper bound on the deflection parameter 
is found to be $d < 2$. 
This result is consistent with our expectation 
$d \leq {\cal O}(1)$. 

Constraints on the parameter $M$ is given by 
consistency of our scenario. 
We have been assuming that $d \neq 0$ is originated from 
the anomaly mediation. 
This point is nothing but 
the crucial difference of our scenario 
from the gauge mediation scenario. 
Therefore, SUSY breaking in the messenger sector 
should be negligible compared with 
the original SUSY breaking in the hidden sector. 
This requirement is described as 
\begin{eqnarray} 
\frac{| \langle W \rangle |}{M_{pl}^2} 
\ll |F_\phi| \sim m_{3/2}  \; ,
\end{eqnarray}  
where $M_{pl}$ and $m_{3/2}$ are the reduced Planck mass 
and the gravitino mass, respectively. 
Using the above solutions, we find 
\begin{eqnarray} 
M \ll \left(
\frac{p (p-1)}{3-p}
\right)^{\frac{- p}{2 (3-p)}} \; 
\left( 
\frac{|F_\phi|}{M_{pl}}
\right)^{\frac{1}{3-p}}  \;  M_{pl} \; .  
\label{constraint}
\end{eqnarray}  
Note that this condition is also consistent 
with a natural requirement $ r \ll M_{pl}$.

\section{A concrete model} 

In the previous section, we have generalized the deflection 
parameter to the positive region. 
The simple example consistent with our assumption 
is the superpotential with negative $p$. 
We can hardly imagine that any perturbative theories 
have such a superpotential, 
the so-called runaway-type superpotential. 
However, there occurs the case in the SUSY gauge theories 
through non-perturbative gauge dynamics 
\cite{Affleck:1983mk}. 
Now we present a concrete model. 

Our model is based on the strong gauge group 
$SU(N_c)$ ($N_c \geq 2$) with the particle contents as follows. 
\begin{center}
\begin{tabular}{ccc}
 \hspace{1cm} &  $~SU(N_c)~$ & $U(1)_R$   \\
$\overline{Q}$ &  $\overline{\bf N}$  & $1-N_c$ \\
$Q$ &  $ {\bf N} $ & $1-N_c$ \\
$Z$ &  $ {\bf 1} $ & $2 N_c$ \\
$S$ &  $ {\bf 1} $ & $1-N_c$ \\
\end{tabular}
\end{center}
The general (renormalizable) superpotential is given by 
\begin{eqnarray} 
W = Z \left( 
\left( \overline{Q} Q \right) - S^2 \right) 
+ (N_c-1) \frac{\Lambda^{\frac{3 N_c-1}{N_c-1}}}
{(\overline{Q}Q)^{\frac{1}{N_c-1}}} \;  , 
\end{eqnarray}  
where the second term is the dynamically 
generated superpotential \cite{Affleck:1983mk}, 
and $\Lambda $ is the dynamical scale. 
We have omitted dimensionless free parameters 
for simplicity. 

After integrating out the superfields 
$\overline{Q}$, $Q$ and $Z$ under 
their SUSY vacuum conditions, 
we obtain the effective superpotential, 
\begin{eqnarray} 
W_{eff} = (N_c-1) \Lambda^{\frac{3 N_c-1}{N_c-1}} 
S^{\frac{2}{1-N_c}} \; ,
\end{eqnarray}  
which corresponds to the superpotential 
of Eq.~(\ref{superpotential}) 
with the identifications,  
$p  = -2/(N_c-1) < 0$ and 
$M = (N_c-1)^{\frac{N_c-1}{3 N_c-1}} \Lambda $. 
The deflection parameter is found to be 
$ d = 2 (N_c-1)/(N_c+1) > 0$. 

For simplicity, let us take a special limit $N_c \gg1 $, 
which leads to $p \sim 0$ and thus $d \sim 2$. 
The condition of Eq.~(\ref{constraint}) 
gives the upper bound on the dynamical scale, 
$\Lambda \ll (F_\phi/N_c M_{pl})^{1/3} M_{pl}$. 
Taking a reasonable value 
$F_\phi \sim {\cal O}(10~\mbox{TeV})$ 
in the AMSB scenario, 
we can find that 
the messenger scale 
$r \sim \sqrt{\Lambda/F_\phi} \Lambda \leq M_{GUT}$ 
and $\Lambda \leq 10^{12}$ GeV 
are consistent with the condition 
with $N_c \sim {\cal O}(10)$, 
where $M_{GUT} \sim 10^{16}$ GeV is the grand unification scale.

\section{ Sparticle mass spectrum} 

Now let us figure out 
the sparticle mass spectrum in our scenario. 
General formulas are given by the method 
developed in Ref.~\cite{Giudice:1997ni} 
(see also \cite{Pomarol:1999ie}). 
For the general case with $F_S/S = d F_\phi$, 
they are found to be 
\begin{eqnarray} 
\frac{m_{\lambda_i}}{\alpha(\mu)} 
&=& \frac{F_\phi}{2} \left(
\frac{\partial}{\partial \mbox{ln} \mu} 
- d \frac{\partial}{\partial \mbox{ln}|S|} 
\right) \alpha^{-1} (\mu, S)  \nonumber \\ 
m^2_i(\mu) &=& 
-\frac{|F_\phi|^2}{4} 
 \left( 
  \frac{\partial}{\partial \mbox{ln} \mu} 
  - d \frac{\partial}{\partial \mbox{ln}|S|} 
   \right)^2 
 \mbox{ln} Z_i(\mu, S)  \; .
\label{softmass}
\end{eqnarray}  
All we have to know is the dependence of 
the gauge coupling $\alpha (\mu, S)$ 
and the wave function $Z_i(\mu,S)$  
on the renormalization scale $\mu$ and 
on the singlet $S$ after integrating out the messengers. 
In the parentheses, the first and the second terms 
correspond to the purely anomaly mediated contribution 
and the additional corrections 
through the messengers, respectively. 
Note that the limit $|d| \gg 1$ reduces 
the formulas to that in the gauge mediation scenario 
\cite{Giudice:1997ni}. 

For a simple gauge group, the gauge coupling and 
the wave functions are given by 
\begin{eqnarray} 
\alpha^{-1} (\mu, S) 
&=&
 \alpha^{-1} (\Lambda_{cut})  
+ \frac{b-N}{4 \pi} 
  \mbox{ln} \left( 
 \frac{S^\dagger S}{\Lambda_{cut}^2} \right)
 + \frac{b}{4 \pi} \mbox{ln} \left( 
 \frac{\mu^2}{S^\dagger S} \right)  \; , 
\label{gaugino} \\ 
Z_i (\mu, S) 
&=& 
Z_i(\Lambda_{cut} ) 
\left(\frac{\alpha(\Lambda_{cut})}{\alpha(S)} 
\right)^{\frac{2 c_i}{b-N}} 
\left(\frac{\alpha(S)}{\alpha(\mu)} 
\right)^{\frac{2 c_i}{b}}   \; , 
 \label{scalar}  
\end{eqnarray}  
where $\Lambda_{cut}$ is the ultraviolet cutoff, 
$b$ is the beta function coefficient,
and $c_i$ is the quadratic Casimir. 
Substituting them into Eq.~(\ref{softmass}), 
we obtain 
\begin{eqnarray} 
m_{\lambda}(\mu ) &=&
 \frac{\alpha(\mu)}{4 \pi} F_\phi (b+ d N)  \; , 
\label{gauginomass}  \\ 
m_i^2 (\mu) &=& 2 c_i \left(
\frac{\alpha(\mu)}{4 \pi} \right)^2
 |F_\phi|^2 \; b \; G(\mu, S) \; ,
\label{scalarmass} 
\end{eqnarray}  
where 
\begin{eqnarray} 
G(\mu, S) = 
\left( \frac{N}{b} \xi^2 + \frac{N^2}{b^2} (1-\xi^2)  
\right) d^2 
+ 2 \frac{N}{b} d  + 1 
\end{eqnarray}  
with 
\begin{eqnarray} 
\xi \equiv   \frac{\alpha(S)}{\alpha(\mu)}
= \left[ 1+ \frac{b}{4 \pi} \alpha(\mu) \mbox{ln} 
\left(
\frac{S^\dagger S}{\mu^2} \right) 
\right]^{-1}   \; . 
\end{eqnarray}   
In $d=0$, Eq.~(\ref{scalarmass}) leads to 
the squared-masses negative 
for an asymptotically non-free gauge theory ($b <0$ ).  
This occurs as the tachyonic slepton problem 
in the minimal supersymmetric standard model (MSSM). 

Let us extract the threshold corrections 
due to the heavy messengers to 
the sparticle squared-masses. 
Taking $\xi=1$ at the messenger scale $\mu=S$  
and subtracting the purely anomaly mediated contribution, 
we obtain
\begin{eqnarray} 
 m_i^2|_{th} &=& 2 c_i \left(
 \frac{\alpha(S)}{4 \pi} \right)^2
 |F_\phi|^2 N \; d (d+2)  \; . 
\label{threshold}
\end{eqnarray}  
We can see that $ m_i^2|_{th} < 0$ 
for the conventional scenario, $(-2 <) d < 0$. 
This is the reason that the scenario is called 
the ``anti-gauge mediation''. 
On the other hand, in our scenario, 
the threshold correction has just the same sign 
as in the gauge mediation scenario. 
For this reason, we may call our scenario 
``anomaly induced gauge mediation''.  
 
It is straightforward to extend the above formulas 
to that for the sparticles in the MSSM.  
Neglecting the effects of Yukawa couplings, 
the sparticle masses (in GeV) 
evaluated at $\mu =100$ GeV are depicted in Fig.~1 
as a function of $\mbox{Log}_{10}[S/\mbox{GeV}]$ 
for the case $d=2$ and $N=2$ with $F_\phi=20$ TeV. 
Here we have taken the gauge couplings 
in the standard model such that 
$\alpha_3(m_Z) \sim 0.12$,  
$\alpha_2(m_Z) \sim 0.033$, 
and $\alpha_1(m_Z) \sim 0.017$. 
We can find that the resultant spectrum is similar to 
that of the gauge mediation scenario. 
However, our result is the distinctive one, 
since the deflection parameter is, at most, 
of order one, and far from the ``gauge mediation limit'', 
$d \gg 1$. 
We also present the result of conventional 
deflected anomaly mediation scenario \cite{Pomarol:1999ie}
in Fig.~2 with $d=-1$ and $N=4$. 
It is interesting to compare these two graphs. 
The opposite sign of the deflection parameter 
causes the big difference. 
Note that we need less number of the messenger fields 
than that in the conventional scenario 
in order to fix the tachyonic slepton problem. 
For example, introduction of only one pair of messengers 
is enough in the case $d=2$ as can be seen in Fig.~3. 

The lightest sparticle is B-ino in Fig.~1, and 
is a candidate of the LSP.  
There is another candidate in the conventional scenario,  
the fermionic partner of $S$. 
Analyzing the scalar potential, we find 
that its mass is of order $F_\phi$  
due to the superpotential. 
Therefore, in our scenario, 
the LSP is always the sparticle in the MSSM. 
Although what is the LSP depends on the parameters, 
$d$ and $N$, and the messenger scale, 
we can find that the LSP is mostly B-ino 
providing the solution of the tachyonic slepton problem. 
This result may be reasonable,  
since existence of charged LSP is usually problematic 
in the cosmological point of view.

\section{Conclusion}

Although the AMSB is very attractive scenario in supergravity,
it cannot be phenomenologically viable 
because of its prediction of the tachyonic slepton. 
It is inevitable that the AMSB scenario should be extended 
in order to fix the problem, 
even if its beautiful feature,  
namely, model independence, is somewhat lost. 

As one elegant scenario, 
we considered the deflected anomaly mediation scenario. 
If there is a sizable deflection parameter, 
the messenger sector plays the essential role 
so that the tachyonic slepton problem can be fixed. 
In the conventional scenario, 
only negative deflection parameter has been taken into account. 
Based on the simple superpotential, 
we generalized the scenario to the case 
with the positive deflection parameter. 
Furthermore, we presented the concrete model 
which could naturally realize this scenario. 

Sparticle masses were found to be 
quite different from that in the conventional scenario, 
but similar to that in the gauge mediation scenario. 
However, it is the distinctive one,  
since the corrections through the messengers 
and the purely AMSB contributions are of the same order. 
This is because the SUSY breaking in the messenger sector 
is originated from the superconformal anomaly. 
This point is the crucial difference 
from the conventional gauge mediation scenario. 
It may be reasonable to call our scenario 
``anomaly induced gauge mediation''. 
 
An elegant mechanism to solve the $\mu$-problem 
was proposed in the original deflected anomaly 
mediation scenario \cite{Pomarol:1999ie}. 
Since the mechanism is independent of 
the sign of the deflection parameter, 
we can follow the same manner, 
and obtain the $\mu$-term of the same order 
as the sparticle masses.

\section*{Acknowledgments}
The author would like to thank Markus Luty 
for useful discussions and comments.


\newpage
\begin{figure}
\begin{center}
\epsfig{file=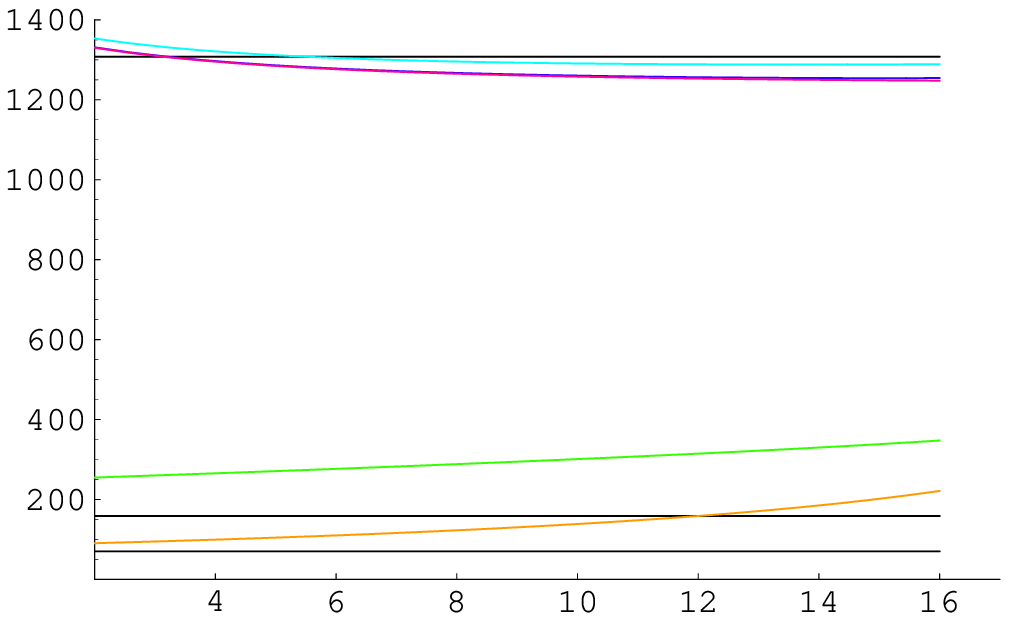, width=12cm}
\caption{
Soft masses 
(absolute values for the gaugino masses) 
of the left-handed squark ($m_{\tilde{Q}} $), 
the right-handed up-squark ($m_{\tilde{U}}$), 
the right-handed down-squark ($m_{\tilde{D}}$), 
the gluino ($m_{\tilde{g}}$), 
the left-handed slepton ($m_{\tilde{L}}$), 
the W-ino ($m_{\tilde{W}}$), 
the right-handed slepton ($m_{\tilde{E}}$) 
and the B-ino ($m_{\tilde{B}}$) 
are plotted from above at the messenger scale 100 GeV. 
Here $d=2$, $N=2$ and $F_\phi =20$ TeV have been taken. 
Two lines of $m_{\tilde{U}}$ and $m_{\tilde{D}}$ 
are almost overlapped, and not distinguishable. 
}
\label{fig:Fig1}
\end{center}
\end{figure}

\newpage
\begin{figure}
\begin{center}
\epsfig{file=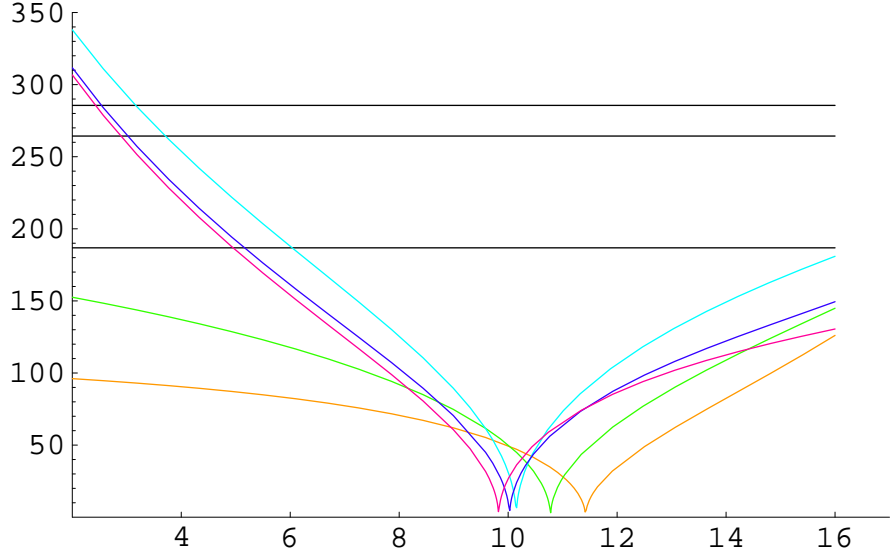, width=12cm}
\caption{
Soft masses (absolute values) 
in the conventional scenario 
\cite{Pomarol:1999ie} with $d=-1$ and $N=4$. 
$|m_{\tilde{Q}}|$, $|m_{\tilde{U}}|$, 
$|m_{\tilde{D}}|$, $|m_{\tilde{B}}|$, 
$|m_{\tilde{W}}|$, $|m_{\tilde{g}}|$, 
$|m_{\tilde{L}}|$ and $|m_{\tilde{E}}|$ 
are plotted from above at the messenger scale 100 GeV. 
The left hand side from the cusps for each graph 
is the negative squared-masses region, 
and is the phenomenologically excluded region. 
} 
\label{fig:Fig2}
\end{center}
\end{figure}

\newpage
\begin{figure}
\begin{center}
\epsfig{file=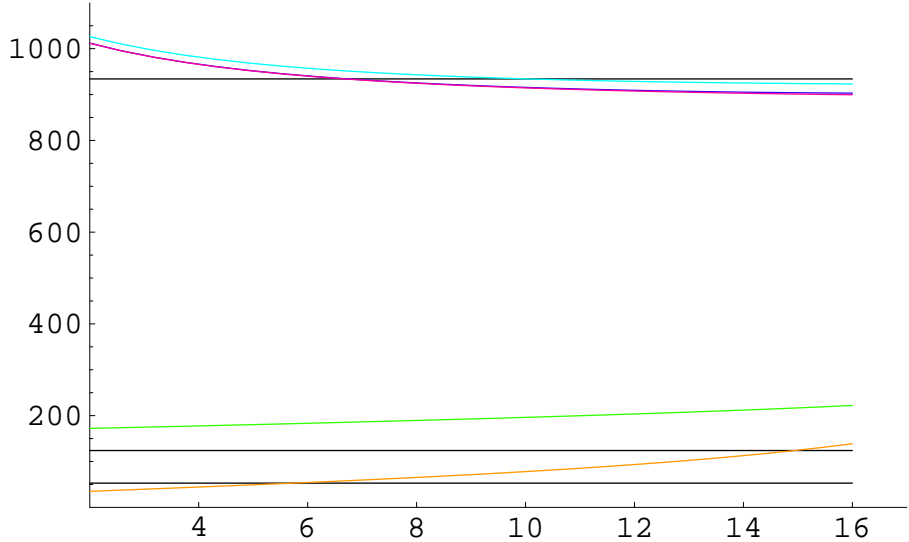, width=12cm}
\caption{
Soft masses for the case $d=2$ and $N=1$ 
with $F_\phi =20$ TeV. 
$m_{\tilde{Q}}$, $m_{\tilde{U}}$, 
$m_{\tilde{D}}$,  $|m_{\tilde{g}}|$, 
$m_{\tilde{L}}$,  $|m_{\tilde{B}}|$, 
$|m_{\tilde{W}}|$ and $m_{\tilde{E}}$ 
are plotted from above at the messenger scale 100 GeV. 
Two lines of $m_{\tilde{U}}$ and $m_{\tilde{D}}$ 
are almost overlapped, and not distinguishable. 
} 
\label{fig:Fig3}
\end{center}
\end{figure}
%
\end{document}